# Extracting the trustworthiest way to service provider in complex online social networks

Lianggui Liu

*School of Information Science and Technology, Zhejiang Sci-Tech University, Hangzhou, 310018, China.*

**Abstract**   In complex online social networks, it is crucial for a service consumer to extract the trustworthiest way to a target service provider from numerous social trust paths between them. The extraction of the trustworthiest way (namely, optimal social trust path (OSTP)) with multiple end-to-end quality of trust (QoT) constraints has been proved to be NP-Complete. Heuristic algorithms with polynomial and pseudo-polynomial-time complexities are often used to deal with this challenging problem. However, existing solutions cannot guarantee the efficiency of searching, that is, they can hardly avoid obtaining partial optimal solutions during searching process. Quantum annealing uses delocalization and tunneling to avoid falling into local minima without sacrificing execution time. It has been proved to be a promising way to many optimization problems in recently published literatures. In this paper, for the first time, QA based OSTP algorithms (QA_OSTP) is applied to the extraction of the trustworthiest way. The experiment results show that QA based algorithms have better performance than its heuristic opponents.

**Keywords**   Quality of trust, trustworthiest way, quantum annealing, quantum tunneling, complex online social networks

## 1  Introduction

Online social networks (OSNs) [1, 2] have gained much attention from the world recently. According to Nielson's report in 2009, more than two-thirds of the global online population visit and participate in online social networks [3]. In such kind of networks, the network structure is made up of participants such as individuals or organizations, and links between participants such as interactions, relationships, and connections. Since there may be over thousands of ways between a pair of interactive participants in large scale OSNs [4], a primary concern arise spontaneously, that is, for a source participant (e.g. a service consumer), how to exactly extract the trustworthiest way to a target participant (e.g. a service provider) within a tolerable period of time? Unfortunately, the extraction of the trustworthiest way, in other words, optimal social trust path (OSTP) selection with multiple end-to-end quality of trust (QoT) constraints has been proved to be NP-Complete [5]. Evaluating the trustworthiness of the target participant along all these social trust paths requires large amount of computation time [6]. Polynomial and pseudo-polynomial-time heuristic algorithms are often used to deal with this problem. However, existing solutions cannot guarantee the quality of final configuration or the speed of searching. Thus this field is attracting more and more attention recently with the rapid development of OSNs.

## 2  Related work

### 2.1 *Multiple QoS Constrained Path Selection*

Korkmaz *et.al.* [7] firstly proposed a approximation algorithm H_MCOP to determine a feasible path that satisfies a set of constraints while maintaining high utilization of network resources. In their algorithm, both multi-constrained values and values of QoS parameters values are aggregated based on (1).

---

*Corresponding author (email: lianggui liu@126.com)

$$g_\lambda(p) \stackrel{def}{=} (\frac{w_1(p)}{c_1})^\lambda + (\frac{w_2(p)}{c_2})^\lambda + ... + (\frac{w_K(p)}{c_K})^\lambda \qquad (1)$$

where $\lambda \geq 1$; $w_i(p)$ is the aggregated value of the $i^{th}$ QoS parameter of path $p$; $c_i$ is the $i^{th}$ QoS constraint value of the selected path between source node $s$ and destination node $d$. H_MCOP firstly use Dijkstra's algorithm to find the path with the minimum $g_\lambda$ from $s$ and $d$, which intends to find whether there exists a feasible solution satisfying all end-to-end QoS parameters in a sub-network. During this process, the aggregated value of each QoS parameter for the identified path from $s$ to $d$ is computed and recorded at each intermediated node along the path. If there exists at least one feasible solution, then these aggregated values are used in another search from $s$ to $d$, which intends to identify a feasible path from $s$ to $d$ with the minimal cost.

Yu *et.al.* [8] studied the problem of service selection with multiple QoS constraints and proposed an approximation algorithm, MCSP-K based on H_MCOP, which checks the number of paths it currently keeps and removes the path with the maximum $g_\lambda$ if the path number reaches $K$. In their algorithm the paths with $K$ minimum $g_\lambda$ will be kept at each intermediate node, which will ensure that MCSP-K will never prune out a feasible path if one exists. In the service candidate graph therein, there is a link between any two nodes in adjacent service sets. If this requirements cannot be satisfied in a network, MCSP_K will search all the paths from source node to each intermediate which will lead the time complexity to be exponential. Thus, this algorithm does not fit for large scale complex online social networks.

## 2.2 *Heuristic algorithm for OSTP*

Till now, there are only a few works that are proposed to address the problem in complex online social networks where some significant influence factors including trust, recommendation roles and social relationships are taken into account.

Liu,G *et.al.* [5] developed a novel efficient heuristic algorithm for OSTP selection named MFPB_HOSTP in complex online social networks based on Dijkstra's algorithm [9]. In MFPB_HOSTP, they first proposed the objective function given in (2)

$$\delta(p) \stackrel{def}{=} max\{(\frac{1-T_p}{1-c_1}),(\frac{1-r_p}{1-c_2}),(\frac{1-\rho_p}{1-c_3})\} \qquad (2)$$

where $T_p$, $r_p$, $\rho_p$ are QoT parameters and represent trust information between participants, social relationships between adjacent participants and recommendation roles of a participant, respectively. They adopted the *Backward_Search* procedure to identify the path with the minimal $\delta$ from $s$ to $d$ to investigate whether there exists a feasible solution where all end-to-end QoT constraints can be satisfied in the sub-network, and to record the aggregated QoT parameters of the path identified from $s$ to each intermediate node. If a feasible solution exists, MFPB_HOSTP then adopts the *Forward_Search* procedure to find a near-optimal path from $s$ to $d$. This process adopts the information provided by *Backward_Search* to identify whether there is another path which satisfies QoT constraints. In this process, MFPB_HOSTP first searches the path with maximal $\mathscr{F}$ value from $s$. Assume one of the adjacent nodes of $s$, $m$ is selected based on Dijkstra's shortest path algorithm as the utility of the path from $s$ to it is maximal. Let $p_{m \to d}^{b(\delta)}$ denote the backward local path from $m$ to $d$ identified in the *Backward_Search* procedure. Then a foreseen path from $s$ to $d$ via $m$ is formed. If this foreseen path is feasible, then MFPB_HOSTP chooses the next node from $m$ with the maximal $\mathscr{F}$ value which is calculated based on Dijkstra's shortest path algorithm. Otherwise, MFPB_HOSTP does not search the path from $m$ and the link $s \to m$ will be deleted from the sub-network. Subsequently, MFPB_HOSTP performs the *Forward_Search* procedure to search the path from $s$ in the sub-network without the link $s \to m$. MFPB_HOSTP is one of the most promising algorithms in solving the OSTP selection problem as it outperforms prior exiting algorithms in both efficiency and the quality of delivered solutions [5].

## 2.3 *Quantum annealing (QA)*

In statistical mechanics, a physical process called simulated annealing (SA) is often performed in order to relax the system to the state with the minimum energy. In the basic form of SA, it first generates an initial solution as the current feasible solution using Metropolis algorithms[10, 11]. Then another solution is selected in the neighborhood of the current solution and replaces the current solution with the new one with the following transition probability given by Metropolis criterion:

$$P(i \Rightarrow j) = \begin{cases} 1 & \text{if } f(j) \leq f(i) \\ \exp(\frac{f(i)-f(j)}{t}) & \text{otherwise} \end{cases} \qquad (3)$$

where $t \in R^+$ represents the control parameter. $f(i)$ and $f(j)$ are energy functions corresponding to state $i$ and $j$ respectively. The same process continues iteratively lots of times. Non-optimal configuration with probability $\exp(\frac{f(i)-f(j)}{t})$ is used to avoid being stuck in a local optimization each time, although the goal is to find a global optimal configuration. Ob-viously, result of one arbitrary taste is only dependent upon the result of the previous taste, thus concepts in a Markov Chain corresponding to a control temperature $t$ can be used. As to SA, one-step transition matrix in a Markov Chain is defined as follows:

$$P_t(i,j) = \Pr[s(q+1) = j | s(q) = i] = \begin{cases} 0 & \text{if } j \notin N(i) \text{ and } j \neq i \\ G(i,j) \min\{1, e^{\frac{f(i)-f(j)}{T}}\} & \text{if } j \in N(i) \text{ and } j \neq i \\ 1 - \sum_{i' \neq i} G(i,i') \min\{1, e^{\frac{f(i)-f(j)}{T}}\} & \text{otherwise} \end{cases} \quad (4)$$

where $G(i,j)$ represent the probability with which configuration $j$ is derived from $i$, and $N(i) \subseteq S$ is the neighborhood set of $i$. With the temperature decreasing, only the better deterioration configuration can be accepted. Simulated annealing is a particularly promising minimization technique. It has, for example, proved effective in finding the global minimum of multidimensional functions having large numbers of local minima [12].

It is worth mentioning that introducing thermal fluctuation is not the only way to perform annealing; QA depends on quantum fluctuation instead [13]. An prominent advantage of quantum fluctuation over thermal fluctuation originates from the fundamental property of the quantum theory, namely, the possibility of tunneling through classically impenetrable potential barriers between energy valleys. Consequently, methods of quantum search, in principle, could be more efficient than the classical search methods [14-17]. Furthermore, the effect of quantum tunneling is shown to be crucial for solving many computationally difficult problems, including the class of nondeterministic polynomial time problems. A practical implementation of QA will need to solve the time-dependent Schrödinger equation in a very large and exponentially growing Hilbert space, which can only rely on a robust quantum computer. Recent research in this area have been carried out by Path-Integral Monte Carlo (PIMC) simulations using quantum-classical mapping with the aid of a Suzuki-Trotter transformation [18, 19], inspired from which we apply QA to the OSTP in complex OSNs. To the best of our knowledge this is the first application of QA to the OSTP and experiment results shows QA based searching algorithm has better performance than its heuristic opponent.

The structure of the present paper is as follows. In Section 3, we describe our proposed algorithm in detail. Specifically, in Section 3.1 Quantum Hamiltonian are given, which is then approximated by a classical one with the aid of a Suzuki–Trotter transformation. The QA based OSTP algorithms (QA_OSTP) is presented and efficiency concerns are addressed in Section 3.2 and 3.3, respectively. Section 4 contains experimental results and analysis which show that QA_OSTP outperforms MFPB_HOSTP. Finally, we concludes this paper in Section 5 with a summary and research prospects.

## 3  Algorithm description

In order to give prominence to the main problem of OSTP in complex online social networks, the *Enron* email dataset [20], a widely used system in the investigation of social networks [5, 21-23] is selected for our research and the related parameter sets such as QoT set, set of weight of QoT parameters *et. al.* are the same as those in [5] for comparison.

In order to solve OSTP in large scale complex online social networks using quantum annealing, we first need to map the problem onto a highly constrained Ising model [24, 25]. Then searching the optimal social trust path is corresponding to finding the ground state with the lowest energy in an Ising model. Moreover, some assumptions should be given first as follows to eliminate some secondary factors which will increase unnecessary complexity and may influence the performance of the complex online social networks:

**Assumption 1**: Before searching the OSTP, three QoT parameters, that is, the values of trust, the social intimacy degree between participants and the role impact factor of participants mentioned in reference [5] have been already obtained through mining techniques [22].

**Assumption 2**: Since the transverse Ising spin glass (TISG) model is the simplest model in which quantum effects in a random system can and have been studied extensively and systematically [26], here we focus only on the TISG.

**Assumption 3:** The complex online networks studied here are *symmetric*. Symmetric means that each pair of adjacent participants, truster and trustee, can reverse their roles without changing their trust values. For example, in Fig. 1, we set $T_{AC} = T_{CA}$, $r_{AC} = r_{CA}$.

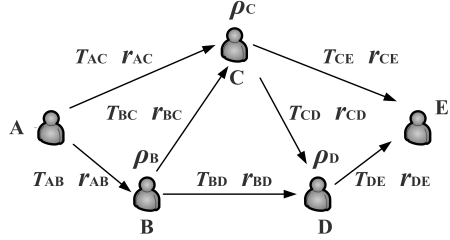

**Figure 1.** Complex online social networks

### 3.1 Problem Representation

In a TISG model, phase space is spanned by all the set of spin variables $\{\omega_\chi\}$, each of which corresponds to a possible configuration.

(1). Quantum Hamiltonian

In QA, the total Hamiltonian for the OSTP can be expressed as a time-dependent Hamiltonian:

$$H_{OSTP} = H_{pot} + H_{kin} \tag{5}$$

where $H_{pot}$ represents the classical potential energy of a given configuration, and $H_{kin}$ is a suitable kinetic energy operator providing the necessary quantum fluctuations to escape local minima. In QA, we seek to minimize $H_{pot}$ as side effect of minimizing $H_{OSTP}$. A suitable configuration is reached if and only if $H_{pot}$ is zero.

In TISG model, the total Hamiltonian in (5) can be rewritten as:

$$H_{OSTP} = H_{TISG} + H_{TF}(t) \tag{6}$$

where $H_{TISG}$ denotes potential energy of TISG model, and $H_{TF}(t)$ is a fictitious kinetic energy introduced typically by the time-dependent transverse field.

A TISG model consists of a set of spins, each of which can only be in one of two states. Each of these spin variable usually takes on the value of either +1 or -1, also known as an up-spin and a down-spin respectively. Formally, for a complex online social networks with $N$ participants each configuration of the system (a feasible social trust path) is associated to a $N \times N$ matrix $U$ with 0/1 entries in the following way: For each pair of the participants $i$ and $j$, if the directed social trust path (an ordered sequence of participants in complex online social networks) goes through the link between $i$ and $j$, then $U_{ij} = 1$, or else $U_{ij} = 0$. Here we renumber the independent variables as $U_k$ $(k=1,...,K)$ and the other dependent variables can be expressed by $U_k$. Then, in quantum mechanics, the quantum Hamiltonian of the OSTP can be expressed as a $K$-spins TISG model through the transformation $U_k \rightarrow (1+\sigma_z^k)/2$, where $\sigma_z^k$ is the Pauli matrix of qubit $k$. And the possible social trust paths can be represented by different quantum states of the $K$ spins.

$$H_{TISG} \equiv \sum_{i=1}^{K} v_i \sigma_z^i + \sum_{i=1}^{K} \sum_{j=i+1}^{K} J_{ij} \sigma_z^i \sigma_z^j \tag{7}$$

We use Fig. 2 and TABLE 1 as a demonstration to describe this kind of correspondence.

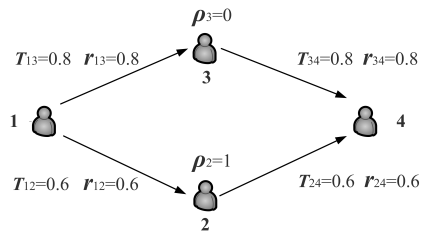

**Figure 2.** An example of complex OSNs
($w_T = 0.3$, $w_r = 0.3$, $w_\rho = 0.4$)

In Fig. 2 a complex OSNs is composed by four participants, where all the parameters are same as those defined in [5]. There are two possible social trust paths between participant 1 and participant 4, that is, 1—3—4 and 1—2—4. The utility values of these two paths determined by the utility function in [5] are 0.384 and 0.744, respectively. Thus even path 1—3—4 has larger $T$ and $r$, participant 1 will choose 1—2—4 as its preferred social trust path according to the weights specified by itself.

**TABLE 1**. Possible social trust path and corresponding state vector in Fig. 2

| $U_{ij}$ | Social trust path | State vector | Utility |
|---|---|---|---|
| $U_{12} = 1$<br>$U_{13} = 0$ | 1—2—4 | $\lvert 01 \rangle$ | 0.744 |
| $U_{12} = 0$<br>$U_{13} = 1$ | 1—3—4 | $\lvert 10 \rangle$ | 0.384 |
| $U_{12} = 1$<br>$U_{13} = 1$ | Invalid path | $\lvert 11 \rangle$ | Invalid |
| $U_{12} = 0$<br>$U_{13} = 0$ | Invalid path | $\lvert 00 \rangle$ | Invalid |

Note that the realization of quantum annealing requires introducing an artificial and adjustable quantum kinetic operator which can provide the quantum fluctuations to escape the local minima. Moreover the quantum annealing process is required to be slow enough to approximating the adiabatic evolution. Reasonably, the choice of $H_{kin}$ should encompass one important question that is determining which configuration are to become direct neighbors of a given configuration. Here we use a effective tactics named "*plus-minus*" [27] as disturbance mechanism to realize a neighborhood of a given configuration. Then in TISG model, to implement QA, a fictitious kinetic energy is introduced typically by the time-dependent transverse field:

$$H_{\text{Kin}} \equiv \Gamma(t) \sum_{i=1}^{K} S^{+}_{\langle i',i \rangle} S^{+}_{\langle j',j \rangle} S^{-}_{\langle j,i \rangle} S^{-}_{\langle j',i \rangle} \tag{8}$$

where $\Gamma(t)$ is the time-dependent power of the transverse field, each $S^{\pm}_{\langle i,j \rangle}$ flips an Ising spin variable at position $(i, j)$ and at the symmetric position $(j,i)$, i.e., $S^{\pm}_{\langle i,j \rangle} = S^{\pm}_{i,j} S^{\pm}_{j,i}$. Then we have the following form of the total quantum Hamiltonian of OSTP:

$$H_{OSTP} = \sum_{i=1}^{K} v_i \sigma_z^i + \sum_{i=1}^{K} \sum_{j=i+1}^{K} J_{ij} \sigma_z^i \sigma_z^j + \Gamma(t) \sum_{i=1}^{K} S^{+}_{\langle i',i \rangle} S^{+}_{\langle j',j \rangle} S^{-}_{\langle j,i \rangle} S^{-}_{\langle j',i \rangle} \tag{9}$$

Initially the strength of the transverse field $\Gamma(t)$ is chosen to be very large, and $H_{OSTP}$ is dominated by the third term of (9). Then $\Gamma(t)$ is gradually and monotonically decreased toward zero, leaving eventually only the first two terms. Note that every state of TISG model can be described as state vector $\lvert \psi(t) \rangle$ and it will evolve with time and should follow the RT Schrödinger equation,

$$i \frac{d}{dt} \lvert \psi(t) \rangle = H(t) \lvert \psi(t) \rangle \tag{10}$$

With $\Gamma(t)$ decreasing, accordingly the state vector $\lvert \psi(t) \rangle$ is expect to evolve from the trivial initial ground state of transverse-field term (6) to the nontrivial ground state of (7), which is the solution of the OSTP.

Then an important issue arises, that is, how slowly we should decrease $\Gamma(t)$ to keep the state vector arbitrarily close to the instantaneous ground state of $H_{OSTP}$. As mentioned in Section 2.3, we will not attempt an actual Schrödinger annealing evolution of the quantum Hamiltonian due to the large Hilbert space. On the contrary, we address the quantum problem by PIMC-QA [25], where annealing will take place in the fictitious time represented by the number of Monte Carlo steps. However, in order to figure out OSTP by PIMC-QA, Suzuki-Trotter transformation should be performed in advance, which requires calculation of the matrix elements of a exponential operator between arbitrary configurations $\lvert \psi \rangle$ and $\langle \psi' \rvert$ of the system, a complicated issue for $H_{\text{Kin}}$. Moreover, since the energy gap between the ground state and the first excited state is large at the beginning, and decreases with the annealing time; hyperbolic interpolation makes the annealing process more efficient and smoother than the linear one does [18], we make the strength of the transverse field as following hyperbolic interpolation:

$$\Gamma(t) = \frac{(1 - t/\eta)\zeta}{t/\eta + \xi} \tag{11}$$

where $\eta$ is the total annealing time and $\zeta, \xi$ are two control operators. This form is trivially Trotter-discretized [18, 19], since the spin-flip term acts independently on single spins at each time slot.

(2). Suzuki–Trotter transformation

In PIMC-QA, a quantum Hamiltonian is approximated by a classical one with the aid of a Suzuki–Trotter transformation. This is possible because of an analogy with a standard TISG model in a transverse field [13]. The transformation maps the quantum Hamiltonian to an effective classical Hamiltonian $H$ similar to the one mentioned in reference [24], and then (9) can be rewritten as:

$$H = \frac{1}{P}\sum_{\rho=1}^{P} H_{\text{TISG}}(\{\Upsilon_{i,\rho}\}) - J_{\Gamma}(\sum_{\rho=1}^{P-1}\sum_{i}\Upsilon_{i,\rho}\Upsilon_{i,\rho+1} + \sum_{i}\Upsilon_{i,1}\Upsilon_{i,P}) \tag{12}$$

where $H$ can be viewed as a consisting of $P$ replicas $\{\Upsilon_{i,\rho}, \rho = 1,...,P\}$ of the classical potential energy of a given configuration $H_{pot}(\{\Upsilon_i\})$, with an interaction of a combined kinetic energy between them, $\Upsilon_{i,\rho}$ denotes the $i$th spin of the $\rho$ th replica. The term $J_{\Gamma}$ is the coupling among the replicas which can be written as:

$$J_{\Gamma} = -\frac{T}{2}\ln\tanh(\frac{\Gamma(t)}{PT}) > 0 \tag{13}$$

where $T$ is the temperature at which each replica is simulated.

### 3.2 Proposed QA_OSTP

In this part, we will describe the components of our algorithms in detail. Notations that are used in QA_OSTP are shown in TABLE 2. Flow chart of QA_OSTP are given in Fig. 3. The first diamond (outermost loop) is controlled by $T$. Here a linear annealing schedule consisting of the initial temperature $T_0$ and $Max_{\text{steps}}$ is selected. Each Monte Carlo step for QA_OSTP consists of a loop starting from the second diamond where $M$ is a tunable multiplier, $M*N$ moves are conducted at each step after which the control parameter is decreased. QA_OSTP keep making the next Monte Carlo Step each time until the termination condition is satisfied. The replicas are always connected to each other in numerical order in the same way throughout the search for the purpose of spin products. The random disturbing means only changing the order in which replicas are selected for search. We find this is a efficient scheme which can promote diversity in the population of configurations.

**TABLE 2** Notations used in QA_OSTP

| | |
|---|---|
| $G$ | Social graph of complex online social networks with multiple QoT parameters |
| $P$ | The number of replicas |
| $T$ | Temperature parameter |
| $T_0$ | Optimal initial temperature |
| $\Gamma_0$ | An initial value of the transverse field |
| $M$ | Fixed number of nearest neighbors of participant |
| $N$ | Number of nodes in complex online social networks |
| $\tau_{\text{total}}$ | Total Monte Carlo time |
| $Max_{\text{steps}}$ | Maximum number of Monte Carlo steps |
| $N_i$ | Number of iterations |
| $\omega$ | A social trust path configuration |
| $\varpi$ | A series of $\omega$, also known as $\{\omega_{\rho}\}$ |
| $\omega_{\rho}'$ | Neighbor of configuration $\omega_{\rho}$ |

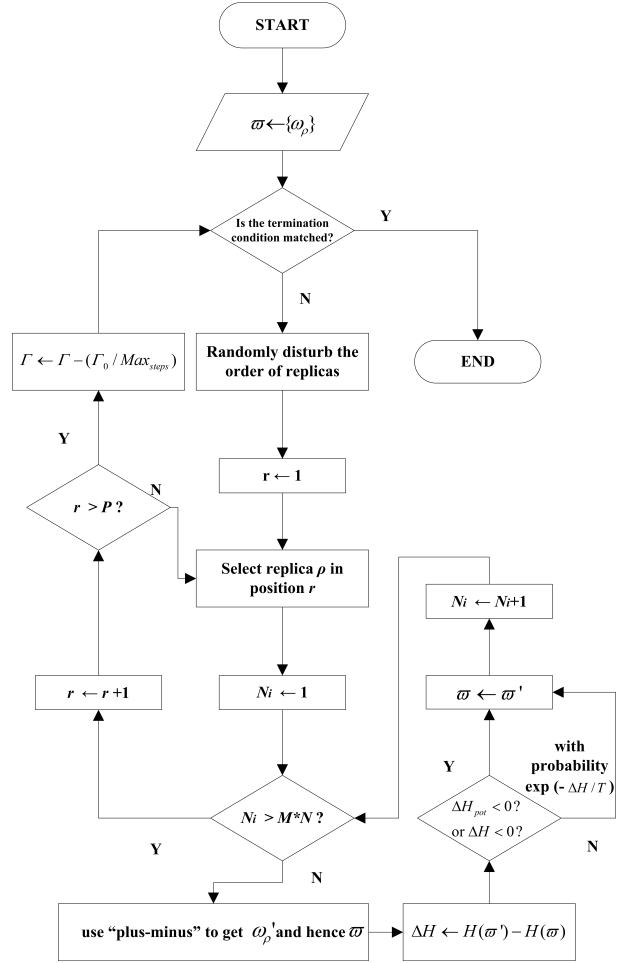

**Figure 3.** Flow chart of QA_OSTP($G,P,T_0,\Gamma_0,Max_{steps}$)

### 3.3 Theoretical analysis

The convergence conditions for the implementation of QA_OSTP with the quantum Monte Carlo evolution is investigated in this part.

In path-integral method, the $d$-dimensional TISG is mapped to a $(d+1)$-dimensional classical Ising system. By using the Suzuki-Trotter transformation, in temperature $T$, when the length of the extra dimension $L \to \infty$, the system partition function can be defined as:

$$Z_T = \sum_{\{\varpi_i^{(k)}\}} \exp\left(\frac{1}{\kappa TL}\sum_{k=1}^{L}\sum_{\langle ij \rangle} J_{ij}\delta_i^{(k)}\delta_j^{(k)} + J_\Gamma \sum_{k=1}^{L}\sum_{i=0}^{N}\delta_i^{(k)}\delta_i^{(k+1)}\right) \tag{14}$$

where $\kappa$ is Boltzmann's constant, $\delta_i^{(k)}$ represents a classical Ising spin at site $i$ on the $k$th Trotter slice, $\varpi$ is the space of discrete configuration.

A Monte Carlo step is characterized by the transition probability from configuration $\omega_\rho$ to configuration $\omega_\rho'$ at time step $\tau$:

$$\Im(\omega_\rho',\omega_\rho;\tau) = P(\omega_\rho',\omega_\rho)\Lambda(\omega_\rho',\omega_\rho;\tau) \tag{15}$$

where $\omega_\rho$ is the present configuration, $\omega_\rho'$ is the next candidate configuration which is generated with defined probability, both of which belong to the space of discrete configuration $\varpi$, $P(\omega_\rho',\omega_\rho)$ is the generation probability with which $\omega_\rho'$ can be generated from $\omega_\rho$. The acceptance probability $\Lambda(\omega_\rho',\omega_\rho;\tau)$ of QA_OSTP can be written as:

$$\Lambda(\omega_\rho',\omega_\rho;\tau) = \Phi\left(\frac{\Theta(\omega_\rho';\tau)}{\Theta(\omega_\rho;\tau)}\right) \tag{16}$$

$$\Theta(\omega_\rho;\tau) = \frac{1}{Z_T}\exp\left(-\frac{C_0(\omega_\rho)}{T_0} - \frac{C_1(\omega_\rho)}{T_1(\tau)}\right) \tag{17}$$

where $C_0(\omega_\rho)$ means the cost function whose global minimum is the optimal social trust path of OSTP problem in OSNs, $C_1(\omega_\rho)$ represents the kinetic energy, $\Theta(\omega_\rho;\tau)$ is the stationary distribution of the homogeneous Markov chain defined by matrix $\Im$ at a given $T_1(\tau)$, $\Phi(\cdot)$ is the monotone increasing acceptance function satisfying $0 \leq \Phi(\cdot) \leq 1$.

To derive the convergence conditions for the implementation of QA_OSTP with the quantum Monte Carlo evolution, we should prove that inhomogeneous Markov chain associated with QA_OSTP are strongly ergodic under appropriate conditions [28]. Our main results are summarized in the following theorems.

**Theorem 1**: For a causal system the transition matrix $\Im$ has the following lower bound:

$$\Im(\omega_\rho',\omega_\rho;\tau) \geq p\Phi(1)\exp\left(-\frac{L_0}{T_0} - \frac{L_1}{T_1(\tau)}\right) \tag{18}$$

where $p$ is the minimum nonvanishing value of $P(\omega_\rho',\omega_\rho)$, $L_0$ and $L_1$ are the maximum changes in a single step in $C_0(\omega_\rho)$ and $C_1(\omega_\rho)$, respectively.

*Proof:* Directly following the definition of the transition probability and the property of the acceptance function in (15), (16) and (17), for both positive $L_0$ and $L_1$, if $\Theta(\omega_\rho';\tau)/\Theta(\omega_\rho;\tau) < 1$, we get

$$\Im(\omega_\rho',\omega_\rho;\tau) \geq p\Phi\left(\frac{\Theta(\omega_\rho;\tau)}{\Theta(\omega_\rho';\tau)}\right)\frac{\Theta(\omega_\rho';\tau)}{\Theta(\omega_\rho;\tau)} \geq p\Phi(1)\exp\left(-\frac{L_0}{T_0} - \frac{L_1}{T_1(\tau)}\right) \tag{19}$$

else we have

$$\Im(\omega_\rho',\omega_\rho;\tau) \geq p\Phi(1) \geq p\Phi(1)\exp\left(-\frac{L_0}{T_0} - \frac{L_1}{T_1(\tau)}\right) \tag{20}$$

Theorem 1 is thereby proved. □

**Theorem 2**: The inhomogeneous Markov chain generated by (16) and (17) is strongly ergodic and converges to the equilibrium state which is corresponding to the term $\exp(-C_0(\omega_\rho)/T_0)$ in (17).

*Proof:* In order to prove strong ergodicity, we refer to the conditions for strong ergodicity [28]. If there exists the transition matrix $\mathbb{N}$ on $\varpi$ such that $\mathbb{N}(\omega_\rho'',\omega_\rho) = \mathbb{N}(\omega_\rho'',\omega_\rho')$ for any $\omega_\rho,\omega_\rho',\omega_\rho'' \in \varpi$, then for a causal system, we have

$$\lim_{\tau \to \infty}\|\Im^{\tau,\varpi} - \mathbb{N}\| = 0 \tag{21}$$

Since the Markov chain is proved to be weakly ergodic in reference [28], we consider for any $s > 0$, $p(\tau,s) = \Im^{\tau,\varpi} p_0$, where $p_0$ belongs to the set of probability distributions on $\omega_\rho \in \varpi$. For a fixed $\omega \in \varpi$, the probability distribution can be written as $p(\omega_\rho'') = \mathbb{N}(\omega_\rho'',\omega)$. Then we have

$$\|p(\tau,\omega) - p\| = \sum_{\omega_\rho'' \in \varpi}\left|\sum_{\omega_\rho \in \varpi}\Im^{\tau,\varpi}(z,\omega_\rho)p_0(\omega_\rho) - \mathbb{N}(\omega_\rho'',\omega)\right| \leq \sum_{\omega_\rho'' \in \varpi}\left|\sum_{\omega_\rho \in \varpi}\left[\Im^{\tau,\varpi}(\omega_\rho'',\omega_\rho) - \mathbb{N}(\omega_\rho'',\omega)\right]p_0(\omega_\rho)\right|$$
$$\leq \sum_{\omega_\rho \in \varpi}\|\Im^{\tau,\varpi} - \mathbb{N}\| = |\varpi|\|\Im^{\tau,\varpi} - \mathbb{N}\| \tag{22}$$

then when $\tau \to \infty$, based on (21), we get

$$\sup\{\|p(\tau,s) - p_\omega\|\} = 0 \tag{23}$$

Therefore, the inhomogeneous Markov chain generated by (16) and (17) is strongly ergodic is thereby proved. □

## 4 Experimental results

### 4.1 Experiment settings

In our experiments, if no otherwise specified, all the related parameters are set following reference [5]. In order to evaluate our proposed algorithms, we compare QA_OSTP with MFPB_HOSTP in terms of two key factors, that is, execution time and the utility of the selected social trust path.

For QA_OSTP, we implement a similar PIMC that was used in reference [25] at a fixed low temperature $T$ (we used $T =$

10/3). The quantum model is mapped onto a classical model with an extra imaginary-time dimension, consisting of P ferromagnetically coupled replicas of the original spin problem, at temperature $PT$ [25] (we used $P = 30$). Since QA requires initial configurations equilibrated at temperature $PT$, an obvious choice is to take $PT = 100$ [25]. Finally, the transverse field $\Gamma$ is annealed hyperbolically in a MC time $\tau$ from an initial value $\Gamma_0 = 300$ to a final value of zero. In QA, we used exclusively "*plus-minus*" tactics, with a static neighborhood pruning [29], which restricts the attempted neighborhood realization by allowing only a fixed number $M$ (we used $M = 20$) of nearest neighbors of participant $j$ to be candidates for $j'$. Our MC step consisted of $M*N$ attempted operations of "*plus-minus*" tactics (for QA, in each of the $P$ replicas). In QA, we averaged the best social trust path utility found over up to 100 independent searches.

In our experiments, the three QoT parameters are randomly generated. The end-to-end QoT constraints specified by a source participant are set as $V^T \geq 0.05$, $V^r \geq 0.001$ and $V^\rho \geq 0.3$, respectively. We first randomly select 80 pairs of source and target participants from the *Enron* email dataset with 87,474 nodes and 30,0511 links. Moreover, following the small world characteristic we set the maximal length of a social trust path 6 hops. Then we number the different network scales from 1 to 25, with number of nodes varying from 50 to 400 and number of links varying from 63 to 2356, respectively.

Both algorithms were implemented in Matlab 7.0 and run on a PC with a 3 GHz Intel processor and 3 GB of RAM with Windows 7.

4.2 *Performance analysis*

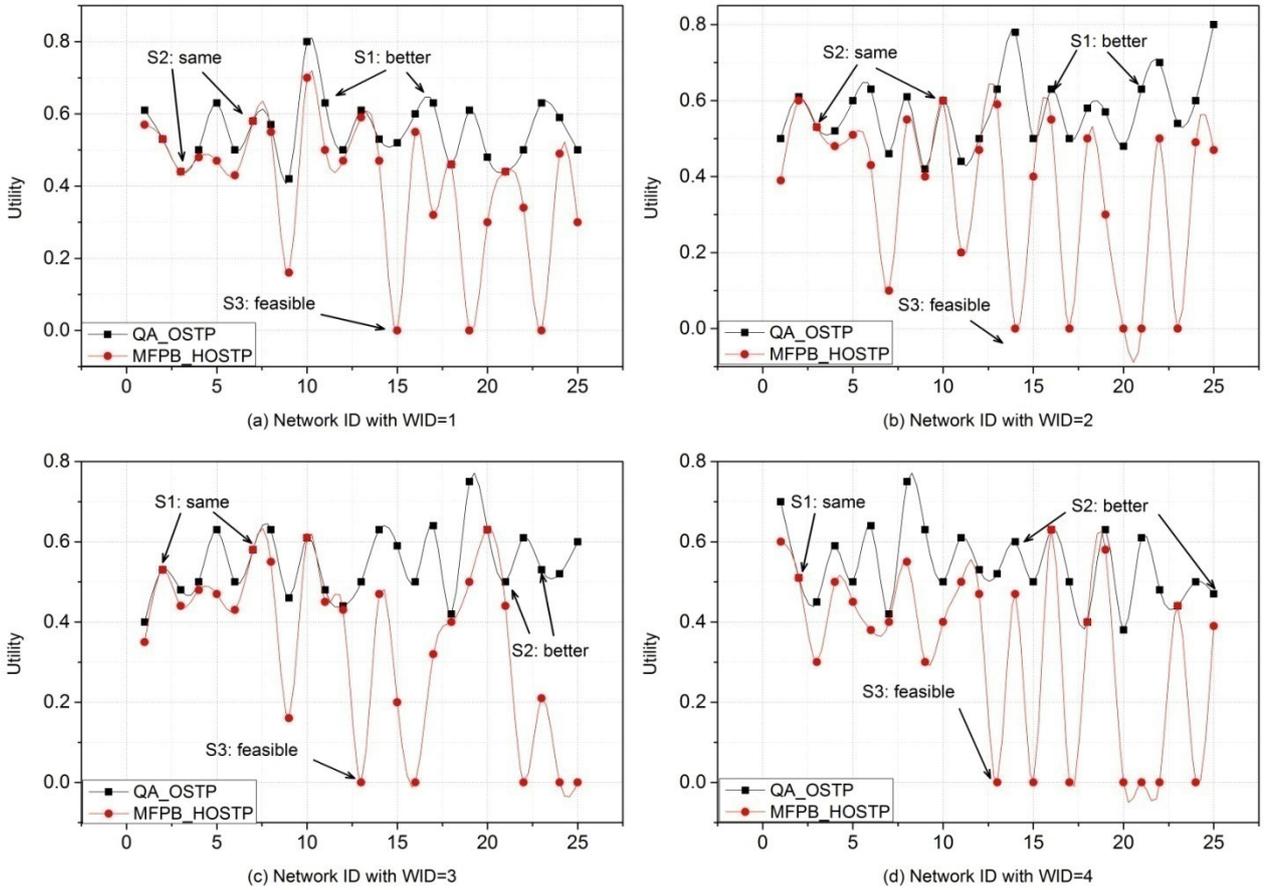

**Figure 4.** Comparison of path utilities of networks

Fig. 4 plots the utilities of the extracted social trust paths with different network scales and different weights of QoT constraints. The weights of QoT parameters are set as $w_T = 0.25, 0.25, 0.5, 0.3$, $w_r = 0.25, 0.5, 0.25, 0.3$, $w_\rho = 0.5, 0.25, 0.25, 0.4$, and each group corresponds to Weight ID 1,2,3,4 respectively. From the figure, we can observe that our QA_OSTP can always find utilities which is not worse than that of MFPB_HOSTP (e.g., case S1 and S2). This is because in QA_OSTP, quantum fluctuation is adopted to avoid the local minima and quantum mechanics works with wave functions that can sample different regions of phase space equally well. While for MFPB_HOSTP, although sub networks is extracted through exhaustive searching before the algorithm execution and *Backward_Search* scheme is used to estimate whether there exists a feasible solution in a sub-network, due to the intrinsic characteristic of heuristic algorithms, it does not guarantee that the best social trust path will be found. Moreover with the network scale growing larger, when the social trust path with the maximal utility is not a feasible solution, the heuristic search can hardly find a near optimal solution and usually returns an infeasible one even when a feasible solution exists (e.g., case S3). Thus, in any case, QA_OSTP shows better performance. In particular, we find that the mean value of utilities of QA_OSTP is 42.85% more than that of MFPB_HOSTP in Fig. 4a, 58.50% more in Fig. 4b, 55.84% more in Fig. 4c, and 63.12% more in Fig. 4d.

Fig. 5 shows the average execution time of algorithms with different weights of QoT parameters and different network scales. Note that, for MFPB_HOSTP, execution time should include the exhaustive searching time for extracting sub-networks, which is not taken into consideration in reference [5]. From Fig. 5, we can see when the network scale is not large, both MFPB_HOSTP and QA_OSTP performs well and the difference is trivial because the searching space is relatively small. But with the network scale expanding, We can observe that QA_OSTP can outperform MFPB_HOSTP in execution time. This is a interesting results since from conventional point, annealing process may be a little time-consuming. The reason we think is that the quantum Hamiltonian in QA_OSTP is approximated with the aid of a Suzuki–Trotter transformation in PIMC-QA and quantum tunneling can avoid some unnecessary searching in MFPB_HOSTP, which can accelerate the annealing process. From Fig. 5, we can see that the average execution time of our proposed algorithm is only 33.47% of that of MFPB_HOSTP in Fig. 5a, 28.36% in Fig. 5b, 31.99% in Fig. 5c and 26.78% in Fig. 5d.

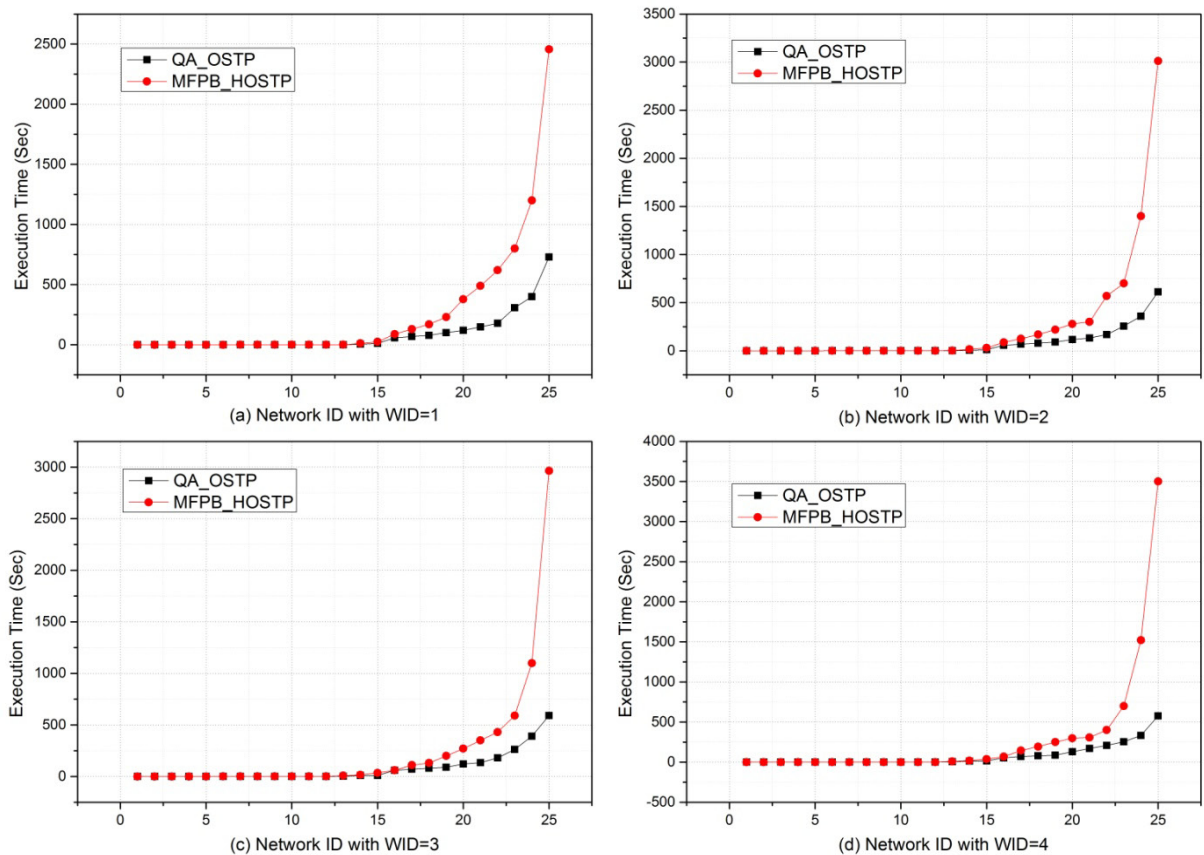

**Figure 5.** Comparison of execution time of algorithms

Based on the above experiments conducted with different scales and parameters, we can observe that QA_OSTP is a promising algorithm and it performs better than its heuristic opponent in both the quality of the selected social trust path and the execution time.

## 5 Conclusion

A novel quantum annealing based OSTP algorithm, that is, QA_OSTP for extracting the trustworthiest way to service provid-er in complex online social networks was proposed. To the best our knowledge, this is the first application of quantum annealing to the challenging NP-Complete OSTP problem in complex online social networks. Due to that quantum mechanics works with wave functions that can sample different regions of phase space equally well, and quantum systems can tunnel through classically impenetrable potential barriers between energy valleys, a process that might prove more effective than waiting for them to be overcome thermally as in SA, QA_OSTP is able to outperform its heuristic opponents and even find configuration of comparable quality to the best algorithms, which shows that QA_OSTP is a very promising tool for solving the OSTP in complex online social networks.

As for the future work, understanding how quantum mechanics can quantitatively improve the quality of solution of OSTP is still a important open issue. Moreover, since the Path-Integral Monte Carlo (PIMC) simulations using quantum-classical mapping with the aid of a Suzuki-Trotter transformation only simulate the equilibrium behavior at finite temperature, we plan to devise another effective and alternative scheme to solve the infinite time Schrödinger equation with stochastic processes.

## Acknowledgements

This work is supported partially by National Natural Science Foundation of China (NSFC) under grant No. 61002016, Zhejiang Provincial Natural Science Foundation of China under grant No. LY13F010016, and Qianjiang Talent Project of Zhejiang Province under Grant No. QJD1302014.